\documentclass[12pt]{article}
\pdfoutput=1
 \usepackage{epsfig}
 \def\be{\begin{equation}}
 \def\ee{\end{equation}}
 \def\bea{\begin{eqnarray}}
 \def\eea{\end{eqnarray}}
 \usepackage{graphicx}

 \catcode`\@=11
 \def\lsim{\mathrel{\mathpalette\@versim<}}
 \def\gsim{\mathrel{\mathpalette\@versim>}}
 \def\@versim#1#2{\vcenter{\offinterlineskip
 \ialign{$\m@th#1\hfil##\hfil$\crcr#2\crcr\sim\crcr } }}
 \catcode`\@=12

 \catcode`@=12
\begin{document}
 \thispagestyle{empty}
 \begin{flushright}
 UCRHEP-T564\\
 March 2016\
 \end{flushright}
 \vspace{0.6in}
 \begin{center}
 {\LARGE \bf Phenomenology of the Utilitarian\\
 Supersymmetric Standard Model\\}
 \vspace{1.2in}
 {\bf Sean Fraser, Corey Kownacki, Ernest Ma, Nicholas Pollard,\\ Oleg 
Popov, and Mohammadreza Zakeri\\}
 \vspace{0.2in}
 {\sl Department of Physics and Astronomy,\\ 
 University of California, Riverside, California 92521, USA\\}
 \end{center}
 \vspace{1.2in}

\begin{abstract}\
We study the 2010 specific version of the 2002 proposed $U(1)_X$ extension 
of the supersymmetric standard model, which has no $\mu$ term and 
conserves baryon number and lepton number separately and automatically.  
We consider in detail the scalar sector as well as the extra $Z_X$ gauge 
boson, and their interactions with the necessary extra color-triplet 
particles of this model, which behave as leptoquarks.  We show how the 
diphoton excess at 750 GeV, recently observed at the LHC, may be explained 
within this context.  We identify a new fermion dark-matter candidate and 
discuss its properties.  An important byproduct of this study is the 
discovery of relaxed supersymmetric constraints on the Higgs boson's mass 
of 125 GeV.
\end{abstract}

 \newpage
 \baselineskip 24pt

\section{Introduction}

Since the recent announcements~\cite{atlas15,cms15} by the ATLAS and CMS 
Collaborations at the Large Hadron Collider (LHC) of a diphoton excess 
around 750 GeV, numerous papers~\cite{many} have appeared explaining 
its presence or discussing its implications.  In this paper, we study the 
phenomenology of a model proposed in 2002~\cite{m02}, which has exactly 
all the necessary and sufficient particles 
and interactions for this purpose. They were of course there for solving 
other issues in particle physics.  However, the observed diphoton excess 
may well be a first revelation~\cite{m16} of this model, including its 
connection to dark matter.

This 2002 model extends the supersymmetric standard model by a new $U(1)_X$ 
gauge symmetry.  It replaces the $\mu$ term with a singlet scalar 
superfield which also couples to heavy color-triplet superfields which 
are electroweak singlets.  The latter are not {\it ad hoc} inventions, 
but are necessary for the cancellation of axial-vector anomalies.  
It was shown in Ref.~\cite{m02} how this was accomplished by the 
remarkable \underline{exact factorization} of the sum of eleven cubic terms, 
resulting in two generic classes of solutions~\cite{m16-1}. 
Both are able to enforce the conservation of baryon number and lepton 
number up to dimension-five terms.  As such, the scalar singlet and 
the vectorlike quarks are indispensible ingredients of this 2002 model. 
They are thus naturally suited for explaining the observed diphoton excess.
In 2010~\cite{m10}, a specific version was discussed, which will be 
the subject of this paper as well.  An important byproduct of this study 
is the discovery of relaxed supersymmetric constraints on the Higgs boson's 
mass of 125 GeV.  This is independent of whether the diphoton excess is 
confirmed or not.

\section{Model}

Consider the gauge group $SU(3)_C \times
SU(2)_L \times U(1)_Y \times U(1)_X$ with the particle content of 
Ref.~\cite{m02}.  For $n_1=0$ and $n_4=1/3$ in Solution (A), the various 
superfields transform as shown in Table 1.  There are three copies of 
$Q,u^c,d^c,L,e^c,N^c,S_1,S_2$; two copies of $U,U^c,S_3$; and one copy of 
$\phi_1,\phi_2,D,D^c$.
\begin{table}[htb]
\caption{Particle content of proposed model.}
\begin{center}
\begin{tabular}{|c|c|c|c|c|}
\hline
Superfield & $SU(3)_C$ & $SU(2)_L$ & $U(1)_Y$ & $U(1)_X$ \\
\hline
$Q = (u,d)$ & 3 & 2 & 1/6 & 0 \\
$u^c$ & $3^*$ & 1 & $-2/3$ & 1/2 \\
$d^c$ & $3^*$ & 1 & 1/3 & 1/2 \\
\hline
$L = (\nu,e)$ & 1 & 2 & $-1/2$ & 1/3 \\
$e^c$ & 1 & 1 & 1 & 1/6 \\
$N^c$ & 1 & 1 & 0 & 1/6 \\  
\hline
$\phi_1$ & 1 & 2 & $-1/2$ & $-1/2$ \\
$\phi_2$ & 1 & 2 & 1/2 & $-1/2$ \\
$S_1$ & 1 & 1 & 0 & $-1/3$ \\
$S_2$ & 1 & 1 & 0 & $-2/3$ \\
$S_3$ & 1 & 1 & 0 & 1 \\
\hline
$U$ & 3 & 1 & 2/3 & $-2/3$ \\
$D$ & 3 & 1 & $-1/3$ & $-2/3$ \\
$U^c$ & $3^*$ & 1 & $-2/3$ & $-1/3$ \\
$D^c$ & $3^*$ & 1 & 1/3 & $-1/3$ \\
\hline
\end{tabular}
\end{center}
\end{table}
The only allowed terms of the superpotential are thus trilinear, i.e.
\begin{eqnarray}
&& Q u^c \phi_2, ~~~ Q d^c \phi_1, ~~~ L e^c \phi_1, ~~~ L N^c \phi_2, ~~~ 
S_3 \phi_1 \phi_2, ~~~ N^c N^c S_1, \\ 
&& S_3 U U^c, ~~~ S_3 D D^c, ~~~ u^c N^c U, ~~~ u^c e^c D, ~~~ d^c N^c D, ~~~ 
Q L D^c, ~~~ S_1 S_2 S_3.
\end{eqnarray}
The absence of any bilinear term means that all masses come from soft 
supersymmetry breaking, thus explaining why the $U(1)_X$ and electroweak 
symmetry breaking scales are not far from that of supersymmetry breaking. 
As $S_{1,2,3}$ acquire nonzero vacuum expectation values (VEVs), the exotic 
$(U,U^c)$ and $(D,D^c)$ fermions obtain Dirac masses from $\langle S_3 
\rangle$, which also generates the $\mu$ term.  The singlet $N^c$ fermion 
gets a large Majorana mass from $\langle S_1 \rangle$, so that the neutrino 
$\nu$ gets a small seesaw mass in the usual way. The singlet $S_{1,2,3}$ 
fermions themselves get Majorana masses from their scalar counterparts 
$\langle S_{1,2,3} \rangle$ through the $S_1 S_2 S_3$ terms.  The only 
massless fields left are the usual quarks and leptons. They then become 
massive as $\phi^0_{1,2}$ acquire VEVs, as in the minimal supersymmetric 
standard model (MSSM).

Because of $U(1)_X$, the structure of the superpotential conserves both 
$B$ and $(-1)^L$, with $B=1/3$ for $Q,U,D$, and $B=-1/3$ for $u^c,d^c,U^c,D^c$; 
$(-1)^L$ odd for $L,e^c,N^c,U,U^c,D,D^c$, and even for all others. Hence 
the exotic $U,U^c,D,D^c$ scalars are leptoquarks and decay into ordinary quarks 
and leptons.  The $R$ parity of the MSSM is defined here in the same way, 
i.e. $R \equiv (-)^{2j+3B+L}$, and is conserved.  Note also that the 
quadrilinear terms $QQQL$ and $u^c u^c d^c e^c$ (allowed in the MSSM) as 
well as $u^c d^c d^c N^c$ are forbidden by $U(1)_X$.  Proton decay is thus 
strongly suppressed.  It may proceed through the quintilinear term 
$QQQL S_1$ as the $S_1$ fields acquire VEVs, but this is a dimension-six 
term in the effective Lagrangian, which is suppressed by two powers 
of a very large mass, say the Planck mass, and may safely be allowed.

\section{Gauge Sector}

The new $Z_X$ gauge boson of this 
model becomes massive through $\langle S_{1,2,3} \rangle = u_{1,2,3}$, whereas 
$\langle \phi^0_{1,2} \rangle = v_{1,2}$ contribute to both $Z$ and $Z_X$. 
The resulting $2 \times 2$ mass-squared matrix is given by~\cite{km97}
\begin{equation}
{\cal M}^2_{Z,Z_X} = \pmatrix{(1/2)g_Z^2(v_1^2+v_2^2) & (1/2)g_Z g_X 
(v_2^2-v_1^2) \cr (1/2)g_Z g_X (v_2^2-v_1^2) & 2g_X^2 [(1/9)u_1^2 + (4/9) u_2^2 + 
u_3^2 + (1/4)(v_1^2 + v_2^2)]}.
\end{equation}
Since precision electroweak measurements require $Z-Z_X$ mixing to be very 
small~\cite{elmp09}, $v_1 = v_2$, i.e. $\tan \beta = 1$, is preferred. 
With the 2012 discovery~\cite{atlas12,cms12} of the 125 GeV 
particle, and identified as the one Higgs boson $h$ responsible for 
electroweak symmetry breaking, $\tan \beta =1$ is not compatible with the 
MSSM, but is perfectly consistent here, as shown already in Ref.~\cite{m10} 
and in more detail in the next section.

Consider the decay of $Z_X$ to the usual quarks and leptons.  Each fermionic 
partial width is given by
\begin{equation}
\Gamma(Z_X \to \bar{f} f) = {g_X^2 M_{Z_X} \over 24 \pi} [c_L^2 + c_R^2],
\end{equation}
where $c_{L,R}$ can be read off under $U(1)_X$ from Table 1.  Thus
\begin{equation}
{\Gamma(Z_X \to \bar{t} t) \over \Gamma(Z_X \to \mu^+ \mu^-)} = 
{\Gamma(Z_X \to \bar{b} b) \over \Gamma(Z_X \to \mu^+ \mu^-)} = {27 \over 5}.
\end{equation}
This will serve to distinguish it from other $Z'$ models~\cite{gm08}.

At the LHC, limits on the mass of any $Z'$ boson depend on its production 
by $u$ and $d$ quarks times its branching fraction to $e^-e^+$ and 
$\mu^-\mu^+$.  In a general analysis of $Z'$ couplings to $u$ and $d$ 
quarks, 
\begin{equation}
{\cal L} = {g' \over 2} Z'_\mu \bar{f} \gamma_\mu (g_V - g_A \gamma_5) f,
\end{equation}
where $f = u,d$.  The $c_u,c_d$ coefficients used in an experimental 
search~\cite{atlas14,cms14} of $Z'$ are then given by
\begin{equation}
c_u = {{g'}^2 \over 2} [(g^u_V)^2 + (g^u_A)^2] B(Z' \to l^-l^+), ~~~ 
c_d = {{g'}^2 \over 2} [(g^d_V)^2 + (g^d_A)^2] B(Z' \to l^-l^+),
\end{equation}
where $l = e,\mu$.  In this model
\begin{equation}
c_u = c_d = {g_X^2 \over 4} B(Z' \to l^-l^+).
\end{equation}
To estimate $B(Z' \to l^-l^+)$, we assume $Z_X$ decays to all SM quarks 
and leptons with effective zero mass, all the scalar leptons with 
effective mass of 500 GeV, all the scalar quarks with effective mass 
of 800 GeV, the exotic $U,D$ fermions with effective mass of 400 GeV 
(needed to explain the diphoton excess), and 
one pseudo-Dirac fermion from combining $\tilde{S}_{1,2}$ (the dark matter 
candidate to be discussed) with mass of 200 GeV.  We find 
$B(Z' \to l^-l^+) = 0.04$, and for $g_X=0.53$, a lower 
bound of 2.85 TeV on $m_{Z_X}$ is obtained from the LHC data based on the 
7 and 8 TeV runs.  

\section{Scalar Sector}

Consider the scalar potential consisting of $\phi_{1,2}$ and $S_{1,2,3}$, where 
only the $S_{1,2,3}$ scalars with VEVs are included. 
The superpotential linking the corresponding superfields is 
\begin{equation}
W = f S_3 \phi_1 \phi_2 + h S_3 S_2 S_1.
\end{equation}
Its contribution to the scalar potential is
\begin{equation}
V_F = f^2 (\Phi_1^\dagger \Phi_1 + \Phi_2^\dagger \Phi_2) S_3^* S_3 
+ h^2 (S_1^* S_1 + S_2^* S_2) S_3^* S_3 + |f \Phi_1^\dagger \Phi_2 + 
h S_1 S_2|^2,
\end{equation}
where $\phi_1$ has been redefined to $\Phi_1 = (\phi_1^+,\phi_1^0)$.  The 
gauge contribution is
\begin{eqnarray}
V_D &=& {1 \over 8} g_2^2 [(\Phi_1^\dagger \Phi_1)^2 + (\Phi_2^\dagger \Phi_2)^2 + 
2(\Phi_1^\dagger \Phi_1)(\Phi_2^\dagger \Phi_2) - 4 (\Phi_1^\dagger \Phi_2) 
(\Phi_2^\dagger \Phi_1)] \nonumber \\ &+& {1 \over 8} g_1^2 
[-(\Phi_1^\dagger \Phi_1) + (\Phi_2^\dagger \Phi_2)]^2 \nonumber \\ 
&+& {1 \over 2} g_X^2 \left[-{1 \over 2} \Phi_1^\dagger \Phi_1 -{1 \over 2} 
\Phi_2^\dagger \Phi_2 - {1 \over 3} S_1^* S_1 - {2 \over 3} S_2^* S_2 
+ S_3^* S_3 \right]^2.
\end{eqnarray}
The soft supersymmetry-breaking terms are
\begin{eqnarray}
V_{soft} &=& \mu_1^2 \Phi_1^\dagger \Phi_1 + \mu_2^2 \Phi_2^\dagger \Phi_2 
+ m_3^2 S_3^* S_3 + m_2^2 S_2^* S_2 + m_1^2 S_1^* S_1 \nonumber \\ 
&+& [m_{12} S_2^* S_1^2 + A_f f S_3 
\Phi_1^\dagger \Phi_2 + A_h h S_3 S_2 S_1 + H.c.].
\end{eqnarray}
In addition, there is an important one-loop contribution from the $t$ quark 
and its supersymmetric scalar partners:
\begin{equation}
V_t = {1 \over 2} \lambda_2 (\Phi_2^\dagger \Phi_2)^2,
\end{equation}
where 
\begin{equation}
\lambda_2  = {6 G_F^2 m_t^4 \over \pi^2} \ln \left( {m_{\tilde{t}_1} 
m_{\tilde{t}_2} \over m_t^2} \right)
\end{equation}
is the well-known correction which allows the Higgs mass to exceed $m_Z$.

Let $\langle \phi^0_{1,2} \rangle = v_{1,2}$ and $\langle S_{1,2,3} \rangle 
= u_{1,2,3}$, we study the conditions for obtaining a minimum of the scalar 
potential $V = V_F + V_D + V_{soft} + V_t$.  We look for the solution 
$v_1=v_2=v$ which implies that
\begin{eqnarray}
\mu_1^2 &=& \mu_2^2 + \lambda_2 v^2 \\
0 &=& \mu_1^2 + A_f f u_3 + f^2 (u_3^2 + v^2) + {1 \over 2} g_X^2 
\left( v^2 + {1 \over 3} u_1^2 + {2 \over 3} u_2^2 - u_3^2 \right) 
+ fh u_1 u_2.
\end{eqnarray}
We then require that this solution does not mix the $Re(\phi_{1,2})$ and 
$Re(S_{1,2,3})$ sectors.  The additional conditions are
\begin{eqnarray}
0 &=& A_f f + (2f^2 - g_X^2) u_3, \\ 
0 &=& {1 \over 3} g_X^2 u_1 + fh u_2, \\ 
0 &=& {2 \over 3} g_X^2 u_2 + fh u_1.
\end{eqnarray}
Hence
\begin{equation}
u_1 = \sqrt{2} u_2, ~~~ fh = {-\sqrt{2} g_X^2 \over 3}.
\end{equation}
The $2 \times 2$ mass-squared matrix spanning $[\sqrt{2} Re(\phi_1^0), 
\sqrt{2} Re(\phi_2^0)]$ is \begin{equation}
{\cal M}^2_\phi = \pmatrix{\kappa + g_X^2 v^2/2 & -\kappa + g_X^2 v^2/2 + 
2 f^2 v^2 \cr -\kappa + g_X^2 v^2/2 + 2 f^2 v^2 & \kappa + g_X^2 v^2/2 + 
2\lambda_2 v^2},
\end{equation}
where
\begin{equation}
\kappa = (2f^2-g_X^2)u_3^2 + {2 \over 3} g_X^2 u_2^2 + 
{1 \over 2} (g_1^2 + g_2^2) v^2.
\end{equation}
For $\lambda_2 v^2 << \kappa$, the Higgs boson $h \simeq 
Re(\phi_1^0+\phi_2^0)$ 
has a mass given by
\begin{equation}
m_h^2 \simeq \left( g_X^2 + 2 f^2 + \lambda_2 \right) v^2,
\end{equation}
whereas its heavy counterpart $H \simeq Re(-\phi_1^0+\phi_2^0)$ has 
a mass given by
\begin{equation}
m_H^2 \simeq (4f^2 - 2g_X^2) u_3^2 + {4 \over 3} g_X^2 u_2^2 + (g_1^2 + g_2^2 
- 2 f^2 + \lambda_2) v^2.
\end{equation}
The conditions for obtaining the minimum of $V$ in the $S_{1,2,3}$ 
directions are
\begin{eqnarray}
0 &=& m_3^2 + g_X^2 u_3^2 + \left( 3h^2 - {4 \over 3} g_X^2 \right) u_2^2 
+ {\sqrt{2} A_h h u_2^2 \over u_3}, \\ 
0 &=& m_2^2 + 2 m_{12} u_2 + \left( 2h^2 + {8 \over 9} g_X^2 \right) u_2^2 + 
\left( h^2 - {2 \over 3} g_X^2 \right) u_3^2 + \sqrt{2} A_h h u_3, \\ 
0 &=& m_1^2 + 2 m_{12} u_2 + \left( h^2 + {4 \over 9} g_X^2 \right) u_2^2 + 
\left( h^2 - {1 \over 3} g_X^2 \right) u_3^2 + {1 \over \sqrt{2}} A_h h u_3.
\end{eqnarray}
The $3 \times 3$ mass-squared matrix spanning $[\sqrt{2} Re(S_1), 
\sqrt{2} Re(S_2), \sqrt{2} Re(S_3)]$ is given by
\begin{eqnarray}
&& m^2_{11} = {4 \over 9} g_X^2 u_2^2 - {1 \over \sqrt{2}} A_h h u_3 
+ {1 \over 3} g_X^2 v^2, ~~~ m^2_{22} = 2 m^2_{11} - 2 m_{12} u_2, \\ 
&& m^2_{12} = m^2_{21} = 2 \sqrt{2} m_{12} u_2 + A_h h u_3 + 2 \sqrt{2} 
\left( h^2 + {2 \over 9} g_X^2 \right) u_2^2 - {\sqrt{2} \over 3} g_X^2 v^2, 
\\ 
&& m^2_{33} = 2 g_X^2 u_3^2 - \sqrt{2} A_h h u_2^2/u_3 + (2f^2-g_X^2)v^2,\\ 
&& m^2_{13} = m^2_{31} = A_h h u_2 + 2 \sqrt{2} \left( h^2 - {1 \over 3} g_X^2 
\right) u_3 u_2,\\
&& m^2_{23} = m^2_{32} = \sqrt{2} A_h h u_2 + 2 \left( h^2 - 
{2 \over 3} g_X^2 \right) u_3 u_2. 
\end{eqnarray}
The $5 \times 5$ mass-squared matrix spanning 
$[\sqrt{2}Im(\phi_1^0),\sqrt{2}Im(\phi_2^0),\sqrt{2}Im(S_1),\sqrt{2}Im(S_2), 
\sqrt{2}Im(S_3)]$ has two zero eigenvalues, corresponding to the 
would-be Goldstone modes
\begin{equation}
(1,1,0,0,0)~~{\rm and}~~(v/2,-v/2,-\sqrt{2}u_2/3,-2u_2/3,u_3),
\end{equation}
for the $Z$ and $Z_X$ gauge bosons.  One exact mass eigenstate is 
$A_{12} = [2 Im(S_1) - \sqrt{2} Im(S_2)]/\sqrt{3}$ with mass given by
\begin{equation}  
m^2_{A_{12}} = -6 m_{12} u_2.
\end{equation}
Assuming that $v^2 << u^2_{2,3}$, the other two mass 
eigenstates are $A \simeq -Im(\phi_1^0)+Im(\phi_2^0)$ and 
$A_S \simeq [u_3 Im(S_1) + \sqrt{2}u_3 Im(S_2) + \sqrt{2} u_2 Im(S_3)]/
\sqrt{u_2^2 + 3u_3^2/2}$ with masses given by 
\begin{eqnarray}
m^2_A &\simeq& (4f^2 - 2g_X^2) u_3^2 + {4 \over 3} g_X^2 u_2^2, \\  
m^2_{A_S} &\simeq& -A_h h \left( {3 u_3 \over \sqrt{2}} + 
{\sqrt{2} u_2^2 \over u_3} \right),
\end{eqnarray}
respectively.  The charged scalar $H^\pm = (-\phi_1^\pm + \phi_2^\pm)/
\sqrt{2}$ has a mass given by
\begin{equation}
m^2_{H^\pm} = (4f^2-2g_X^2) u_3^2 + {4 \over 3} g_X^2 u_2^2 + 
(g_2^2-2f^2) v^2.
\end{equation}

\section{Physical Scalars and Pseudoscalars} 

In the MSSM without radiative corrections,
\begin{eqnarray}
m^2_{H^\pm} &=& m_A^2 + m_W^2, \\ 
m^2_{h,H} &=& {1 \over 2} \left( m_A^2 + m_Z^2 \mp \sqrt{(m_A^2+m_Z^2)^2 - 
4m_Z^2 m_A^2 \cos^2 2 \beta} \right),
\end{eqnarray}
where $\tan \beta = v_2/v_1$.  For $v_1=v_2$ as in this model, $m_h$ 
would be zero.  There is of course the important radiative correction 
from Eq.~(14), but that alone will not reach 125 GeV.  Hence the MSSM 
requires both large $\tan \beta$ and large radiative correction, but 
a significant tension remains in accommodating all data.  In this model, 
as Eq.~(23) shows, $m_h^2 \simeq (g_X^2 + 2f^2 + \lambda_2)v^2$, where $v=123$ 
GeV. This is a very interesting and important result, allowing the Higgs boson 
mass to be determined by the gauge $U(1)_X$ coupling $g_X$ in addition to 
the Yukawa coupling $f$ which replaces the $\mu$ parameter, i.e. $\mu = 
f u_3$.  There is no tension between $m_h = 125$ GeV and the superparticle 
mass spectrum.  Since $\lambda_2 \simeq 0.25$ for $\tilde{m}_t \simeq 1$ TeV, 
we have the important constraint 
\begin{equation}
\sqrt{g_X^2 + 2f^2} \simeq 0.885.
\end{equation}
For illustration, we have already chosen $g_X = 0.53$.  Hence $f=0.5$ and 
for $u_3 = 2$ TeV, $f u_3 = 1$ TeV is the value of the $\mu$ parameter 
of the MSSM.  Let us choose $u_2 = 4$ TeV, then $m_{Z_X} = 2.87$ TeV, 
which is slightly above the present experimental lower bound of 2.85 TeV 
using $g_X = 0.53$ discussed earlier.

As for the heavy Higgs doublet, the four components $(H^\pm,H,A)$ are all 
degenerate in mass, i.e. $m^2 \simeq (4f^2-2g_X^2)u_3^2 + (4/3) g_X^2 u_2^2$ 
up to $v^2$ corrections.  Each mass is then about 2.78 TeV.  In more detail, 
as shown in Eq.~(37), $m_{H^\pm}^2$ is corrected by $g_2^2 v^2 = m_W^2$ plus 
a term due to $f$. As shown in Eq.~(24), $m_H^2$ is corrected by 
$(g_1^2 + g_2^2)v^2 = m_Z^2$ plus a term due to $f$ and $\lambda_2$.  
These are exactly in accordance with Eqs.~(38) and (39).

In the $S_{1,2,3}$ sector, the three physical scalars are mixtures of all 
three $Re(S_i)$ components, whereas the physical pseudoscalar $A_{12}$ has 
no $Im(S_3)$ component.  Since only $S_3$ couples to $U U^c$, $D D^c$, and 
$\phi_1 \phi_2$, a candidate for the 750 GeV diphoton resonance must have 
an $S_3$ component.  It could be one of the three scalars or the 
pseudoscalar $A_S$, or the other $S_3$ without VEV.  In the following, 
we will consider the last option, specifically a pseudoscalar $\chi$ with 
a significant component of this other $S_3$.  This allows the $\chi U U^c$, 
$\chi D D^c$ and $\chi \phi_1 \phi_2$ couplings to be independent of the 
masses of $U$, $D$, and the charged higgsino.  The other scalars and 
pseudoscalars are assumed to be much heavier, and yet to be discovered.

\section{Diphoton Excess}

In this model, other than the addition of $N^c$ for seesaw neutrino masses, 
the only new particles are $U, U^c, D, D^c$ and $S_{1,2,3}$, which are 
exactly the ingredients needed to explain the diphoton excess at the LHC. 
The allowed $S_3 U U^c$ and $S_3 D D^c$ couplings enable the one-loop gluon 
production of $S_3$ in analogy to that of $h$.  
\begin{figure}[htb]
\vspace*{-3cm}
\hspace*{-3cm}
\includegraphics[scale=1.0]{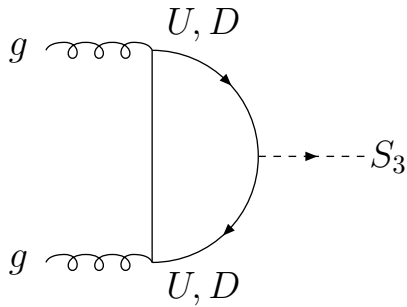}
\vspace*{-21.5cm}
\caption{One-loop production of $S_3$ by gluon fusion.}
\end{figure}
The one-loop decay of 
$S_3$ to two photons comes from these couplings as well as 
$S_3 \phi_1 \phi_2$. 
\begin{figure}[htb]
\vspace*{-3cm}
\hspace*{-3cm}
\includegraphics[scale=1.0]{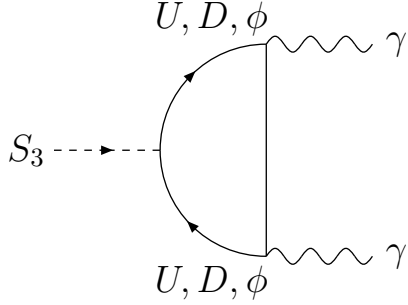}
\vspace*{-21.5cm}
\caption{One-loop decay of $S_3$ to two photons.}
\end{figure}
In addition, the direct $S_1 S_2 S_3$ couplings 
enable the decay of $S_3$ to other final states, including those of the 
dark sector, which contribute to its total width.  The fact that the 
exotic $U,U^c,D,D^c$ scalars are leptoquarks is also very useful for 
understanding~\cite{bn15} other possible LHC flavor anomalies.  In a 
nutshell, a desirable comprehensive picture of possible new physics 
beyond the standard model is encapsulated by this existing model.
In the following, we assume that the pseudoscalar $\chi$ is the 
750 GeV particle, and show how its production and decay are consistent 
with the present data.

The production cross section through gluon fusion is given by
\begin{equation}
\hat{\sigma} (gg \to \chi) = {\pi^2 \over 8 m_\chi^2} \Gamma (\chi \to gg) 
\delta(\hat{s}-m_\chi^2).
\end{equation}
For the LHC at 13 TeV, the diphoton cross section is roughly~\cite{eeqsy15}
\begin{equation}
\sigma(gg \to \chi \to \gamma \gamma) \simeq (100~{\rm pb}) \times 
(\lambda_g~{\rm TeV})^2 \times B(\chi \to \gamma \gamma),
\end{equation}
where $\lambda_g$ is the effective coupling of $\chi$ to two gluons, 
normalized by
\begin{equation}
\Gamma(\chi \to gg) = {\lambda_g^2 \over 8 \pi} m_\chi^3.
\end{equation}
Let the $\chi \bar{Q} Q$ coupling be $f_Q$, then 
\begin{equation}
\lambda_g = {\alpha_s \over \pi m_\chi}~\sum_Q f_Q F(m_Q^2/m_\chi^2),
\end{equation}
where~\cite{hkkt16} 
\begin{equation}
F(x) = 2 \sqrt{x} \left[ \arctan \left( {1 \over \sqrt{4x-1}} \right) \right]^2,
\end{equation}
which has the maximum value of $\pi^2/4 = 2.47$ as $x \to 1/4$.  Let 
$f_Q^2/4 \pi = 0.21$ and $F(m_Q^2/m_\chi^2) = 2.0$ (i.e. $m_Q = 380$ GeV) 
for all $Q = U, U, D$, then $\lambda_g = 0.49$ TeV$^{-1}$.  For the 
corresponding
\begin{equation}
\Gamma(\chi \to \gamma \gamma) = {\lambda_\gamma^2 \over 64 \pi} m_\chi^3,
\end{equation}
the $\phi^\pm$ higgsino contributes as well as $U,D$.  However, its mass 
is roughly $f u_3 = 1$ TeV, so $F(x_\phi) =  0.394$, and
\begin{equation}
\lambda_\gamma = {2 \alpha \over \pi m_\chi}~\sum_\psi N_\psi Q_\psi^2 f_\psi 
F(x_\psi),
\end{equation}
where $\psi = U, U, D, \phi^\pm$ and $N_\psi$ is the number of copies of 
$\psi$.  Using $f_\phi^2/4 \pi = 0.21$ as well, $\lambda_\gamma = 0.069$ 
TeV$^{-1}$ is obtained.  We then have $\Gamma(\chi \to \gamma \gamma) 
= 10$ MeV and $\Gamma(\chi \to g g) = 4.0$ GeV.
If $B(\chi \to \gamma \gamma) = 2.5 \times 10^{-4}$, then $\sigma = 6$ fb, 
and the total width of $\chi$ is 40 GeV, in good agreement with 
data~\cite{atlas15,cms15}.

As mentioned earlier, there are 2 copies of $S_3$ and 3 copies each of 
$S_{1,2}$.  In addition to the ones with VEVs in their scalar components, 
there are 5 other superfields.  One pair $\tilde{S}_{1,2}$ may form a 
pseudo-Dirac fermion, and be the lightest particle with 
odd $R$ parity.  It will couple to $\chi$, say with strength $f_S$ which 
is independent of all other couplings that we have discussed, then 
the tree-level decay $\chi \to \tilde{S}_1 \tilde{S}_2$ dominates the total 
width of $\chi$ and is invisible.
\begin{equation}
\Gamma (\chi \to \tilde{S}_1 \tilde{S}_2) = {f_S^2 \over 8 \pi} 
\sqrt{m_\chi^2 - 4m_S^2}.
\end{equation}
For $m_\chi = 750$ GeV and $m_S = 200$ GeV, we find $\Gamma = 36$ GeV if 
$f_S = 1.2$.  These numbers reinforce our numerical analysis to support 
the claim that $\chi$ is a possible candidate for the 750 GeV diphoton 
excess.  Note also that $\lambda_g$ and $\lambda_\gamma$ have scalar 
contributions which we have not considered.  Adding them will allow us 
to reduce the fermion contributions we have assumed and still get the 
same final reuslts.

If we disregard the decay to dark matter ($f_S=0$), then the total width 
of $\chi$ is dominated by $\Gamma(\chi \to gg)$, which is then less than 
a GeV.  Assuming that the cross section for the diphoton resonance 
is $6.2 \pm 1$ fb~\cite{eeqsy15}, we plot the allowed values of $f^2_Q/4 \pi$ 
versus $m_Q$ for both $f_S = 1.2$ which gives a total width of about 
40 GeV for $\chi$, and $f_S = 0$ which requires much smaller values of 
$f^2_Q/4\pi$.  Since $\chi$ must also decay into two gluons, we show 
the diject exclusion upper limits ($\sim 2$ pb) from the 8 TeV data 
in each case as well.

\newpage
\begin{figure}[htb]
\includegraphics[scale=1.2]{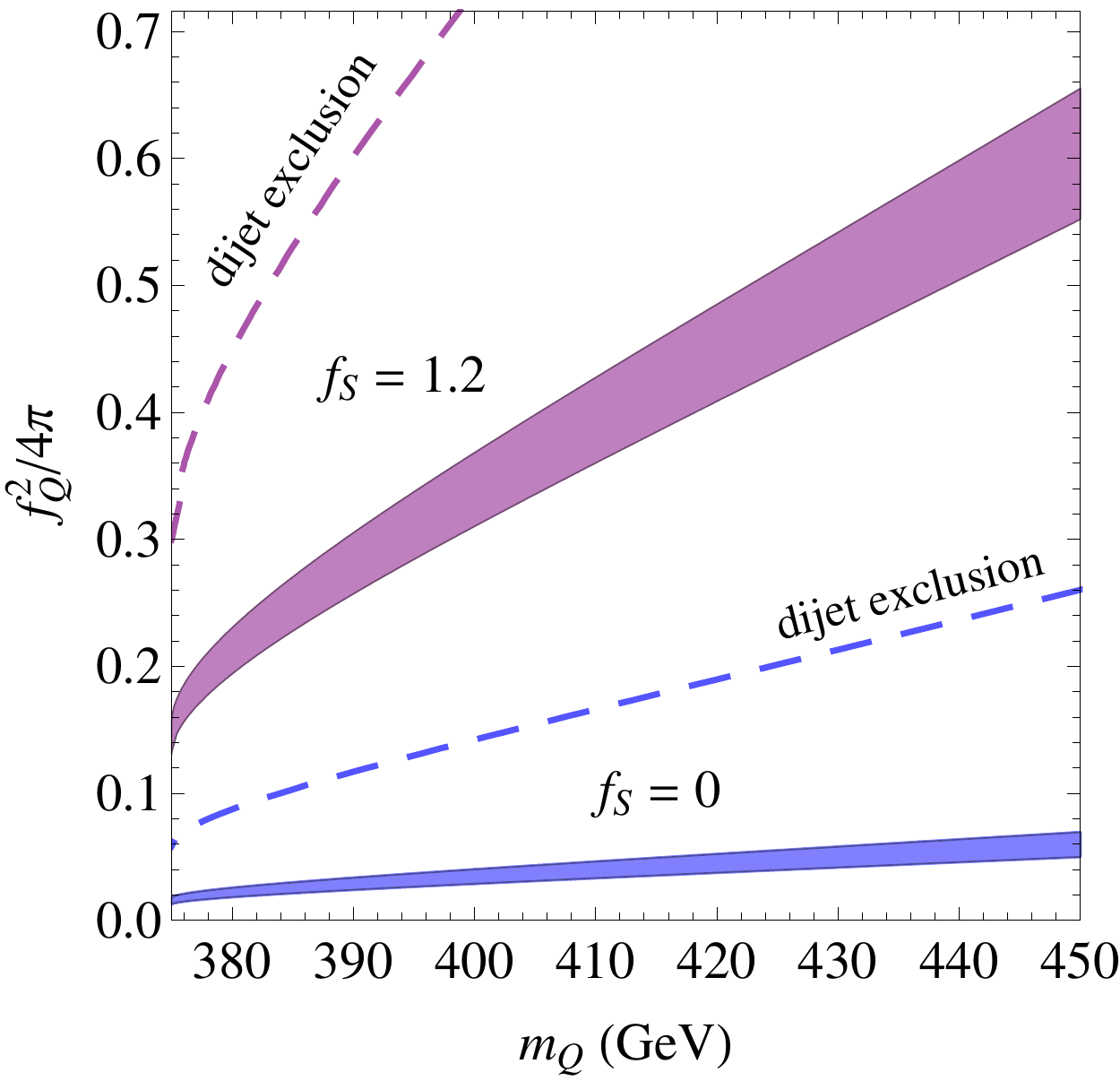}
\caption{Allowed region for diphoton cross section of $6.2 \pm 1$ fb.}
\end{figure}

\section{Scalar Neutrino and Neutralino Sectors}

In the neutrino sector, the $2 \times 2$ mass matrix spanning $(\nu,N^c)$ 
per family is given by the well-known seesaw structure:
\begin{equation}
{\cal M}_\nu = \pmatrix{0 & m_D \cr m_D & m_N},
\end{equation}
where $m_D$ comes from $v_2$ and $m_N$ from $u_1$.
There are two neutral complex scalars with odd $R$ parity per family, i.e. 
$\tilde{\nu} = (\tilde{\nu}_R + i \tilde{\nu}_I)/\sqrt{2}$ and 
$\tilde{N}^c = (\tilde{N}^c_R + i \tilde{N}^c_I)/\sqrt{2}$.  The $4 \times 4$ 
mass-squared matrix spanning $(\tilde{\nu}_R, \tilde{\nu}_I, \tilde{N}^c_R, 
\tilde{N}^c_I)$ is given by
\begin{equation}
{\cal M}^2_{\tilde{\nu},\tilde{N}^c} = \pmatrix{ m^2_{\tilde{\nu}} & 0 & A_D m_D & 0 \cr 
0 & m^2_{\tilde{\nu}} & 0 & -A_D m_D \cr A_D m_D & 0 & m^2_{\tilde{N}^c} + A_N m_N 
& 0 \cr 0 & -A_D m_D & 0 & m^2_{\tilde{N}^c} - A_N m_N}.
\end{equation}
In the MSSM, $\tilde{\nu}$ is ruled out as a dark-matter candidate because 
it interacts elastically with nuclei through the $Z$ boson.  Here, the $A_N$ 
term allows a mass splitting between the real and imaginary parts of the 
scalar fields, and avoids this elastic-scattering constraint by virtue 
of kinematics.  However, we still assume their masses to be heavier than 
that of $\tilde{S}_{1,2}$, discussed in the previous section.

In the neutralino sector, in addition to the $4 \times 4$ mass matrix of 
the MSSM spanning 
$(\tilde{B}, \tilde{W}_3, \tilde{\phi}_1^0, \tilde{\phi}_2^0)$ with the 
$\mu$ parameter replaced by $f u_3$, i.e.
\begin{equation}
{\cal M}_0 = \pmatrix{M_1 & 0 & -g_1 v_1/\sqrt{2} & g_1 v_2/\sqrt{2} \cr 
0 & M_2 & g_2 v_1/\sqrt{2} & -g_2 v_2/\sqrt{2} \cr -g_1 v_1/\sqrt{2} & 
g_2 v_1/\sqrt{2} & 0 & -f u_3 \cr g_1 v_2/\sqrt{2} & -g_2 v_2/\sqrt{2} & 
-f u_3 & 0},
\end{equation}
there is also the $4 \times 4$ mass matrix spanning $(\tilde{X}, \tilde{S}_3, 
\tilde{S}_2, \tilde{S}_1)$, i.e.
\begin{equation}
{\cal M}_S = \pmatrix{M_X & \sqrt{2} g_X u_3 & -2\sqrt{2} g_X u_2/3 & 
-\sqrt{2} g_X u_1/3 \cr \sqrt{2} g_X u_3 & 0 & h u_1 & h u_2 \cr 
-2\sqrt{2} g_X u_2/3 & h u_1 & 0 & h u_3 \cr -\sqrt{2} g_X u_1/3 & h u_2 
& h u_3 & 0}.
\end{equation}
The two are connected through the $4 \times 4$ matrix 
\begin{equation}
{\cal M}_{0S} = \pmatrix{0 & 0 & 0 & 0 \cr 0 & 0 & 0 & 0 \cr -g_x v_1/\sqrt{2} 
& -f v_2 & 0 & 0 \cr -g_X v_2/\sqrt{2} & -f v_1 & 0 & 0}.
\end{equation}
These neutral fermions are odd under $R$ parity and the 
lightest could in principle be a dark-matter candidate.  To avoid the 
stringent bounds on dark matter with the MSSM alone, we assume again that all 
these particles are heavier than $\tilde{S}_{1,2}$, as the dark matter 
discussed in the previous section. 

\section{Dark Matter}

The $5 \times 5$ mass matrix spanning the 5 singlet fermions 
$(\tilde{S}_1,\tilde{S_2},\tilde{S_1},\tilde{S_2},\tilde{S_3})$, 
corresponding to superfields with zero VEV for their scalar components, 
is given by
\begin{equation}
{\cal M}_{\tilde{S}} = \pmatrix{0 & m_0 & 0 & 0 & m_{13} \cr m_0 & 0 & 0 & 0 & 
m_{23} \cr 0 & 0 & 0 & M_3 & M_2 \cr 0 & 0 & M_3 & 0 & M_1 \cr m_{13} & 
m_{23} & M_2 & M_1 & 0}.
\end{equation}
Note that the $4 \times 4$ submatrix spanning $(\tilde{S}_1,\tilde{S_2},
\tilde{S_1},\tilde{S_2})$ has been diagonalized to form two Dirac fermions. 
We can choose $m_0$ to be small, say 200 GeV, and $M_{1,2,3}$ to be large, 
of order TeV.  However, because of the mixing terms $m_{13},m_{23}$, 
the light Dirac fermion gets split into two Majorana fermions, so it 
should be called a pseudo-Dirac fermion.

The dark matter with odd $R$ parity is the lighter of the two Majorana 
fermions, call it $\tilde{S}$, contained in the pseudo-Dirac fermion 
formed out of $\tilde{S}_{1,2}$ as discussed in Sec.~6.  It couples to the 
$Z_X$ gauge boson, but in the 
nonrelativistic limit, its elastic scattering cross section with nuclei 
through $Z_X$ vanishes because it is Majorana.  It also does not couple 
directly to the Higgs boson $h$, so its direct detection at underground 
search experiments is very much suppressed.  However, it does couple to 
$A_S$ which couples also to quarks through the very small mixing of $A_S$ 
with $A$.  This is further suppressed because it contributes only 
to the spin-dependent cross section.  To obtain a spin-independent cross 
section at tree level, the constraint of Eqs.~(17) to (19) have to be 
relaxed so that $h$ mixes with $S_{1,2,3}$.  

Let the coupling of $h$ to $\tilde{S} \tilde{S}$ be $\epsilon$, then 
the effective interaction for elastic scattering of $\tilde{S}$ with 
nuclei through $h$ is given by
\begin{equation}
{\cal L}_{eff} = {\epsilon f_q \over m_h^2} \overline{\tilde{S}} \tilde{S} 
\bar{q} q,
\end{equation}
where $f_q = m_q/2v = m_q/(246~{\rm GeV})$.  The spin-independent 
direct-detection cross section per nucleon is given by
\begin{equation}
\sigma^{SI} = {4 \mu^2_{DM} \over \pi A^2} [\lambda_p Z + (A-Z) \lambda_n]^2,
\end{equation}
where $\mu_{DM} = m_{DM} M_A/(m_{DM} + M_A)$ is the reduced mass of the dark 
matter.  Using~\cite{bbps09}
\begin{equation}
\lambda_N = \left[ \sum_{u,d,s} f_q^N + {2 \over 27} \left( 1 - \sum_{u,d,s} 
f_q^N \right) \right] {\epsilon m_N \over (246~{\rm GeV}) m_h^2},
\end{equation}
with~\cite{jlqcd08}
\begin{eqnarray}
&& f_u^p = 0.023, ~~~ f_d^p = 0.032, ~~~ f_s^p = 0.020, \\ 
&& f_u^n = 0.017, ~~~ f_d^n = 0.041, ~~~ f_s^n = 0.020,
\end{eqnarray}
we find $\lambda_p \simeq 3.50 \times 10^{-8}$ GeV$^{-2}$, and 
$\lambda_n \simeq 3.57 \times 10^{-8}$ GeV$^{-2}$.  
Using $A=131$, $Z=54$, and $M_A = 130.9$ atomic mass 
units for the LUX experiment~\cite{lux15}, and $m_{DM} = 200$ GeV, we 
find for the upper limit of $\sigma^{SI} < 1.5 \times 10^{-45}$ cm$^2$, 
the bound $\epsilon < 6.5 \times 10^{-4}$.

We have already invoked the $\chi \tilde{S}_1 \tilde{S}_2$ coupling to 
obtain a large invisible width for $\chi$.  Consider now the fermion 
counterpart of $\chi$, call it $\tilde{S}'$, and the scalar counterparts 
of $\tilde{S}_{1,2}$, then the couplings $\tilde{S}' \tilde{S}_1 S_2$ and 
$\tilde{S}' \tilde{S}_2 S_1$ are also $f_S = 1.2$.  Suppose one linear 
combination of $S_{1,2}$ , call it $\zeta$, is lighter than 200 GeV, 
then the thermal relic abundance of dark matter is determined by the 
annihilation $\tilde{S} \tilde{S} \to \zeta \zeta$, with a cross 
section times relative velocity given by
\begin{equation}
\sigma \times v_{rel} = {f^4_{\zeta} m_{S'}^2 \sqrt{1 - m^2_{\zeta}/m_S^2} 
\over 16 \pi (m_{S'}^2 + m_S^2 - m_{\zeta}^2)^2}.
\end{equation}
Setting this equal to the optimal value~\cite{sdb12} of 
$2.2 \times 10^{-26}$ cm$^3$/s, 
we find $f_{\zeta} \simeq 0.62$ for $m_{S'} = 1$ TeV, $m_S = 200$ GeV, and 
$m_{\zeta} = 150$ GeV.   Note that $\zeta$ stays in thermal equilibrium 
through its interaction with $h$ from a term in $V_D$.   It is also very 
difficult to be produced at the LHC, because it is an SM singlet, so its 
mass of 150 GeV is allowed.

\section{Conclusion}

The utilitarian supersymmetric $U(1)_X$ gauge extension of the Standard Model 
of particle interactions proposed 14 years ago~\cite{m02} allows for two 
classes of anomaly-free models which have no $\mu$ term and conserve baryon 
number and lepton number automatically.  A simple version~\cite{m10} with 
leptoquark superfields is especially interesting because of existing LHC 
flavor anomalies.

The new $Z_X$ gauge boson of this model has specified couplings to quarks and 
leptons which are distinct from other gauge extensions and may be tested at 
the LHC.  On the other hand, a hint may already be discovered with the 
recent announcements by ATLAS and CMS of a diphoton excess at around 750 GeV.  
It may well be the revelation of the singlet scalar (or pseudoscalar) $S_3$ 
predicted by this 
model which also predicts that there should be singlet leptoquarks and 
other particles that $S_3$ must couple to.  Consequently, gluon fusion 
will produce $S_3$ which will then decay to two photons together with 
other particles, including those of the dark sector.  This scenario 
explains the observed diphoton excess, all within the context of the  
original model, and not an invention after the fact.

Since $S_3$ couples to leptoquarks, the $S_3 \to l_i^+ l_j^-$ decay must 
occur at some level.  As such, $S_3 \to e^+ \mu^-$ would be a very 
distinct signature at the LHC.  Its branching fraction depends on unknown 
Yukawa couplings which need not be very small.  Similarly, the $S_3$ 
couplings to $\phi_1 \phi_2$ as well as leptoquarks imply decays to $ZZ$ and 
$Z \gamma$ with rates comparable to $\gamma \gamma$. 

An important byproduct of this study is the discovery of relaxed 
supersymmetric constraints on the Higgs boson's mass of 125 GeV.  
It is now given by Eq.~(23), i.e. $m_h^2 \simeq (g_X^2 + 2f^2 + \lambda_2)v^2$, 
which allows it to be free of the tension encountered in the MSSM. 
This prediction is independent of whether the diphoton excess is 
confirmed or not.

Most importantly, since $S_3$ replaces the $\mu$ parameter, its identification 
with the 750 GeV excess implies the existence of supersymmetry.  If confirmed 
and supported by subsequent data, it may even be considered in retrospect 
as the first evidence for the long-sought existence of supersymmetry.

\bigskip

\noindent \underline{\it Acknowledgement}~:~
This work was supported in part by the U.~S.~Department of Energy Grant 
No. DE-SC0008541.

\bibliographystyle{unsrt}

\end{document}